\documentstyle [12pt] {article}
\begin{document}

\def\a{{\cal A}}
\def\b{{\cal B}}
\def\c{{\cal C}}
\def\cbar{\underline{\cal C}}
\def\e{{\cal E}}
\def\f{{\cal F}}
\def\kk{{_{K\bar{K}}}}
\def\ppi{{\pi\bar{\pi}}}
\def\kpi{{K\bar{\pi}}}
\def\bkpi{{\bar{K}\pi}}
\def\be{\begin{eqnarray}}
\def\en{\end{eqnarray}}
\def\non{\nonumber}
\def\la{\langle}
\def\ra{\rangle}
\def\pr{{\sl Phys. Rev.}~}
\def\prl{{\sl Phys. Rev. Lett.}~}
\def\pl{{\sl Phys. Lett.}~}
\def\np{{\sl Nucl. Phys.}~}
\def\zp{{\sl Z. Phys.}~}

\font\el=cmbx10 scaled \magstep2
{\obeylines
\hfill ITP-SB-93-49
\hfill UCD-93-31}

\vskip 0.5 cm

\centerline {{\el SU(3) Breaking Effects in Charmed Meson Decays}}
\medskip
\bigskip
\medskip

\centerline{\bf Ling-Lie Chau}
\medskip
\centerline{Department of Physics, University of California}
\centerline{Davis, California 95616, USA}
\medskip
\centerline{and}
\medskip
\centerline{\bf Hai-Yang Cheng}
\medskip
\centerline{ Institute of Physics, Academia Sinica}
\centerline{Taipei, Taiwan 11529, Republic of China}

\centerline{and}

\medskip
\centerline{ Institute for Theoretical Physics, State University of
New York}
\centerline{Stony Brook, New York 11794, USA}

\bigskip
\bigskip

\centerline{\bf Abstract}
\bigskip
   The decay rates of $D^+\to\pi^+\pi^0$ and $D^0\to K^+\pi^-$ recently
measured by CLEO give the ratios $R_1=2|V_{cs}/V_{cd}|^2\Gamma(D^+\to
\pi^+\pi^0)/\Gamma(D^+\to\bar{K}^0
\pi^+)=3.29\pm 1.16$ and $R_2=|V_{cs}^*V_{ud}/(V_{cd}^*V_{us})|^2\Gamma(D^0\to
 K^+\pi^-)/\Gamma(D^0\to K^-\pi^+)=2.92\pm 1.34\,$. Both
are about three times of those expected from SU(3) symmetry.
We show that, in the large-$N_c$ factorization approach, such large SU(3)
violations can be accounted for by the accumulations of several small
SU(3)-breaking effects.  An important requirement is
the relative magnitude of the form factors, $f_+^{D\pi}(0)>f_+^{DK}(0)$.

\pagebreak

    For a long time, theorists [1-3] had observed that, if SU(3) is a good
symmetry, the ratios
\be
R_1 = 2\left|{V_{cs}\over V_{cd}}\right|^2{\Gamma(D^+\to \pi^+\pi^0)\over
\Gamma(D^+\to\bar{K}^0\pi^+)}
\en
and
\be
R_2 = \left|{V_{cs}^*V_{ud}\over V_{cd}^*V_{us}}\right|^2{\Gamma(D^0\to K^+
\pi^-)\over \Gamma(D^0\to K^-\pi^+)}
\en
both ought to be unity. Especially $R_1$ being unity, as had been emphasized
all along, is a clean SU(3) prediction, since both channels are exotic and
there should be no final-state-interaction effects to alter the SU(3) results.
 Quite strikingly, recent data of CLEO
\be
Br(D^+\to\pi^+\pi^0) &=& (0.22\pm 0.05\pm 0.05)\%~~~[4], \\
Br(D^0\to K^+\pi^-) &=& (0.77\pm 0.25\pm 0.25)\%\times Br(D^0\to K^-\pi^+
)~~~[5],
\en
give the values of these ratios
\be
R_1 &=& 3.29\pm 1.16\,,  \\
R_2 &=& 2.92\pm 0.95\pm 0.95\,,
\en
far exeeding unity; a clear large violation of SU(3) results. [ In obtaining
Eqs.(5) and (6), we have used the branching ratio $Br(D^+\to\bar{K}^0\pi^+)=
(2.6\pm 0.4)\%$ given by the Particle Data Group (PDG) [6] and
$(V_{cd}/V_{cs})^2=0.05138\,$.]

  We show in this paper that, the larger-than-unity values of
$R_1$ and $R_2$ can be accounted for in the
   the large $N_c$ factorization approach; the
net large SU(3)-symmetry violations in $R_1$ and $R_2$
are the cumulative results of several small SU(3)-breaking effects [7].

   Let us consider first the decay modes $D^+\to\pi^+\pi^0$ and
$D^+\to\bar{K}^0\pi^+$ in terms of the quark-diagram amplitudes [2,3],
\be
A(D^+\to\pi^+\pi^0) &=& -{G_F\over 2}V_{cd}^*V_{ud}(\a+\b)_\ppi e^{i\delta
_2^\ppi},   \\
A(D^+\to \bar{K}^0\pi^+) &=& {G_F\over\sqrt{2}}V_{cs}^*V_{ud}(\a+\b)_\bkpi
e^{i\delta _{3/2}^\bkpi},
\en
where $\a$ is the external $W$-emission amplitude and $\b$ is the internal
$W$-emission amplitude. These quark-diagram amplitudes  have well-defined
meaning with all QCD strong-interaction effects included. The final-state
interactions are expressed by the phase shifts, the $\delta$'s, which in
general have both real and imaginary parts; the real parts are related to
the elastic scattering effects while the imaginary parts indicate effects of
inelasticity [4]. Since both $\pi^+\pi^0$ and $\bar{K}^0\pi^+$ channels
are exotic, we expect that the phase shifts
$\delta^\ppi_2$ and $\delta^\bkpi_{3/2}$ in Eqs.(7) and (8) are
negligible (actually all we have used in this analysis is that they are real).
If SU(3) symmetry is good, i.e.
$(\a+\b)_\ppi=(\a+\b)_\bkpi$, it is easily seen that $R_1=|\a+\b|^2_\ppi\Big/
|\a+\b|^2_\bkpi=1$ by
substituting Eqs.(7) and (8) into Eq.(1). That was the result given in
Refs.[1,2] more than a decade ago, (in Ref.[1] the invariant SU(3) amplitudes
were used; in Ref.[2] the quark diagram scheme was introduced  and used as
presented here.)

   We can determine the absolute values of the quark diagram amplitude
$|\a+\b|$ in Eqs.(7) and (8) from the aforementioned branching ratios for
$D^+\to\pi^+\pi^0$ and $D^+\to\bar{K}^0\pi^+$:
\be
|\a+\b|_\ppi &=& (0.287\pm 0.065)\,{\rm GeV}^3, \\
|\a+\b|_\bkpi  &=& (0.164\pm 0.018)\,{\rm GeV}^3,
\en
where we have used the total width $\Gamma(D^+)=6.174\times 10^{-13}\,$ GeV [6]
to covert the branching ratios into decay rates.
Clearly they are far from being equal, thus violating SU(3) symmetry.

 In the following we try to understand this large SU(3)-symmetry violation
 in these quark-diagram amplitudes from the state-of-the-art large $N_c$
 factorization approach (for a
review, see [8]). The quark-diagram amplitudes are given by
\be
\a_\ppi &=& -\sqrt{2}a_1(p_{\pi^+})_\mu f_\pi\la\pi^0|\bar{c}\gamma^\mu(1-
\gamma_5)d|D^+\ra,  \non \\
\b_\ppi &=& a_2(p_{\pi^0})_\mu f_\pi\la\pi^+|\bar{c}\gamma^
\mu(1-\gamma_5)u|D^+\ra,   \\
\a_\bkpi &=& a_1(p_{\pi^+})_\mu f_\pi\la\bar{K}^0|\bar{c}\gamma^\mu(1-\gamma_5)
s|D^+\ra,  \non \\
\b_\bkpi &=& a_2(p_{_{\bar{K}^0}})_\mu f_\pi\la\pi^+|\bar{c}\gamma^
\mu(1-\gamma_5)u|D^+\ra,  \non
\en
where
\be
a_1 &=& {1\over 2}(c_++c_-)+{1\over 2N_c}(c_+-c_-),  \non \\
a_2 &=& {1\over 2}(c_+-c_-)+{1\over 2N_c}(c_++c_-),
\en
$c_+$ and $c_-$ are the Wilson coefficient functions to be evaluated at
the renormalization scale $\mu\sim m_c$, and $N_c$ is the number of quark color
degrees of freedom. The two-body matrix
elements in Eq.(11) are in general expressed in terms of the form factors $f_0$
and $f_1$ [9]; for example,
\be
\la\pi^0|\bar{c}\gamma_\mu d|D^+\ra = -{1\over\sqrt{2}}\left\{\left(
(p_D+p_\pi)_\mu-{m^2_D-m^2_\pi\over q^2}q_\mu\right)f_1^{D\pi}(q^2)+{m^2_D-
m^2_\pi\over q^2}q_\mu f_0^{D\pi}(q^2)\right\},
\en
where $q=p_D-p_\pi$ and the factor of $-{1\over\sqrt{2}}$ comes from the wave
function of the $\pi^0$. Substituting Eqs.(11) and (13) into Eqs.(7) and (8)
we obtain
\be
(\a+\b)_\ppi &=& (m^2_D-m^2_\pi)f_\pi f_0^{
D\pi}(m_\pi^2)(a_1+a_2), \non  \\
(\a+\b)_\bkpi &=& a_1(m_D^2-m^2_K)f_\pi f_0^{DK}(m^2_\pi)
+a_2(m^2_D-m^2_\pi)f_Kf_0^{D\pi}(m^2_K).
\en
As noted before, it is reasonable to
assume that $\delta_2^\ppi\sim 0$ and $\delta_{3/2}^{\bar{K}\pi}\sim 0$
since these are exotic channels. Hence, we find from Eqs.(1), (7), (8) and
(14) that
\be
R_1=\,1.073\left( {[m^2_D-m^2_\pi]\over[m^2_D-m^2_K]}\,{f_0^{D\pi}(m^2_\pi)
\over f_0^{DK}(m^2_\pi)}\,{[1+(a_2/ a_1)]\over [1+(a_2/ a_1)r]}\right)^2,
\en
where the factor of 1.073 comes from the phase-space differences for the
$\pi\bar{\pi}$ and $\bar{K}\pi$ modes, and
\be
r=\,{f_K\over f_\pi}\,{m^2_D-m^2_\pi\over m^2_D-m^2_K}\,{f_0^{D\pi}(m^2_K)
\over f_0^{DK}(m^2_\pi)}.
\en

   In order to calculate $(\a+\b)$ and $R_1$, we need information on the form
factors and $c_1,~c_2$.
 The $q^2$ dependence of the form factor $f_0(q^2)$ is usually assumed to be
governed by a single low-lying pole:
\be
f_0(q^2)=\,{f_0(0)\over 1-(q^2/m^2_*)},
\en
where $m_*$ is the mass of the $0^+$ pole. For the form factor $f_0^{DK}(0)
$ at $q^2=0$, we use the average value (recall that $f_0(0)=f_1(0)=f_+(0)
$; see [9])
\be
f_0^{DK}(0)=\,0.76\pm 0.02
\en
extracted from the recent measurements of $D\to K\ell\bar{\nu}$ by CLEO II,
E687 and E691 [10]. For the form factor $f_0^{D\pi}(0)$, there are various
sources of information: A previous measurement
of the Cabibbo-suppressed decay $D^0\to \pi^-\ell^+\nu$ by Mark III [11,10]
yields
\be
\left|{f_0^{D\pi}(0)\over f_0^{DK}(0)}\right|=\,1.0^{+0.6}_{-0.3}\pm 0.1\,;
\en
a very recent CLEO-II measurement of $D^+\to \pi^0\ell^+\nu$ [12] gives
\be
\left|{f_0^{D\pi}(0)\over f_0^{DK}(0)}\right|=\,1.29\pm 0.21\pm 0.11\,,
\en
which clearly indicates that $f_0^{D\pi}(0)$ is most likely greater than
$f_0^{DK}(0)$, in agreement with the current theoretically favored value
$f_0^{D\pi}(0)/f_0^{DK}(0)=1.18$
given by the heavy quark symmetry and chiral perturbation theory [13] and
ruling out  the previous Wirbel-Stech-Bauer results $f_0^{D\pi}(0)=0.69$
 and $f_0^{DK}(0)=0.76$ [14]. For
definiteness, we use $f_0^{D\pi}(0)=0.83$ determined from the experimental
result Eq.(3) for $D^+\to\pi^+\pi^0$. Moreover, we use $f_\pi=132$
MeV and $f_K=161$ MeV [6].

   It is known from analyzing the data of charm decays using the quark diagram
scheme [3,15] that charm decays involving the internal
 $W$-emission diagram are not suppressed, though some earlier
model-calculations had such suppressions [16] and
gave birth to the so-called
 "color-suppression" rule. The cleanest example of such decay is
 $D^0\to\bar{K}^0\phi$ which involves only the  internal $W$-emission diagram.
 The so-called QCD-sum-rule calculation of Blok and Shifman [16] gave too
small branching ratio for it. As far as we know,
such defect in their QCD-sum-rule calculations has not being corrected
 for charmed meson decays. Future measurement of  $D^+\to{K}^+\phi$,
another decay involving only the internal $W$-emission diagram, will further
clarify this point. (See Ref.[15] for detailed discussions on this point and
 critics on Blok and Shifman's QCD-sum-rule calculations for charmed
meson decays).

The absence of suppression on the internal $W$-emission diagrams in charmed
meson decays, together with the destructive interference in the decay of
$D^+\to\bar{K}^0\pi^+$,  was later identified in the large-$N_c$
factorization model calculation [17] to the dropping of the $1/N_c$ terms
(for a review, see [8]). Taught by these experiences, we will neglect
the $1/N_c$ correction in our calculations here.

 Given $c_+(m_c)\cong
0.75$, $c_-(m_c)\cong 1.77$ [18], and neglecting the $1/N_c$ term, we obtain
from Eq.(12)
\be
a_1(m_c)={1\over 2}(c_++c_-)\cong 1.26\,,~~~a_2(m_c)={1\over 2}(c_+-c_-)\cong
-0.51\,.
\en
Using $m_*=2.47$ GeV for $D\to\pi$ transition  and $m_*=2.60$
GeV for $D\to K$, [14], we finally arrive at
\be
R_1=3.3\,,
\en
in excellent agreement with the experimental result Eq.(5).

The ratio $R_1$ is very
sensitive to the relative magnitude of the form factors $f_0^{D\pi}(0)$
and $f_0^{DK}(0)$. For example, if the previous Wirbel-Stech-Bauer results
$f_0^{D\pi}(0)=0.69$ and $f_0^{DK}(0)=0.76$
 [14] were used in our calculation, we would have obtained $R_1=1.4\,$,
in disagreement with data. So the  value of the form factor ratio
needed in our analysis is consistent with the experimental value of
Eq.(20). We further note in
Eq.(15) that the ratio of $\b/\a$ is given by $a_2/
a_1\cong -0.40$ for $D^+\to\pi^+\pi^0$ decay and by $(a_2/a_1)r
\cong -0.60$ for $D^+\to\bar{K}^0\pi^+$ decay; i.e. the destructive
interference in the latter decay is much more severe than in the former, a
SU(3)-symmetry violation effect crucial to  making the value of $R_{1}$
much greater than unity.

 To summerize, we see from Eqs.(15) and (16) that all
the SU(3)-symmetry breaking effects, though small individually,  enhance
cumulatively
 the decay rate of $D^+\to\pi^+
\pi^0$ in comparsion to that of $D^+\to\bar{K}^0\pi^+$: $(m^2_D-m^2_\pi)/(
m^2_D-m^2_K)=1.07\,$, $f_0^{D\pi}(m^2_\pi)/f_0^{DK}(m^2_\pi)=1.092\,$, $f_0^{D
\pi}(m^2_K)/f_0^{DK}(m^2_\pi)=1.134\,$, $f_K/f_\pi=1.22\,$;
each of them gives a SU(3)-symmetry
effect below $25\%$, but the accumulation of them leads to $[1+(c_2/c_1)]
/[1+(c_2/c_1)r]=1.5$ and finally a value of $R_1=3.3\,$.

   We next discuss the decays $D^0\to K^+\pi^-$ and $D^0\to K^-\pi^+$.
In terms of the quark diagram amplitudes [2,3]
\be
A(D^0\to K^-\pi^+)=\,{1\over 3}\,{G_F\over\sqrt{2}}V_{cs}^*V_{ud}
\left[(2\a-\b+3\c)_\bkpi e^{i\delta^
\bkpi_{1/2}}+(\a+\b)_\bkpi e^{i\delta^\bkpi_{3/2}}\right],
\en
where $\c$ is the $W$-exchange amplitude. The expression for $A(D^0\to K^+
\pi^-)$ is exactly the same as Eq.(23) except replacing $V^*_{cs}
V_{ud}$ by $V^*_{cd}V_{us}$ and the subscript and superscript $\bkpi$ by
$\kpi$. It is obvious that if the square brackets in Eq.(23) for $\bkpi$ and
$\kpi$ are the same, then $R_2=1$.

 To calculate $R_2$ we need to have know the phase shift difference
$\Delta_\bkpi=\delta_{1/2}^\bkpi-\delta_{3/2}^
\bkpi$, i.e. the final-state-interaction effects. For the $\bar{K}\pi$ channel
both the the phase shift difference $\Delta_\bkpi$ and the quark-diagram
amplitudes $|\a+\b|$, $|\b-\c|$ in Eq.(23) can be obtained
 from the available data of $D^+\to \bar{K}^0\pi^+,~D^0\to K^-\pi^+$
 and $\bar{K}^0\pi^0$. Using the PDG values for
the decay rates of
$D^+\to \bar{K}^0\pi^+,~D^0\to\bar{K}^0\pi^0$ [6] and the updated result for
the branching ratio of $D^0\to K^-\pi^+$ [19],
$Br(D^0\to K^-\pi^+)=(3.90\pm 0.16)\%$,
which is the weighted average of previous measurements and the new CLEO
result of $(3.95\pm 0.08\pm 0.17)\%$ [19], we obtain (for simplicity,
$\delta_{1/2}^\bkpi$ and $\delta_{3/2}^\bkpi$ are assumed to be real)
\be
|\b-\c|_\bkpi=(0.205\pm 0.010)\,{\rm GeV}^3,~~~\Delta_{\bar{K}\pi}=(90\pm 11)
^\circ,
\en
with $|\a+\b|_\bkpi$ being given by Eq.(10). Unfortunately, we cannot do a
similar detailed quark-diagram analysis for $D^0\to K^+\pi^-$ to obtain
$|\b-\c|_\kpi$ and the phase shift difference
$\Delta_\kpi=\delta_{1/2}^\kpi-\delta_{3/2}^
\kpi$
 directly from experiment in comparison with those of $\bkpi$, since other
quark-mixing doubly-suppressed decays have not been measured. Since we do not
know how to calculate
phase shifts, we shall assume $\Delta_\kpi=\Delta_\bkpi$ and calculate the
amplitudes and see if
we can obtain the SU(3) violation effects in $R_2$ totally from the amplitude.
Indeed we find that we can.

   The quark-diagram amplitudes in
the large $N_c$ factorization approach are given by:
\be
\a_{\bar{K}\pi} &=& a_1(m_D^2-m_K^2)f_\pi f_0^{DK}(m^2_\pi),  \non \\
\b_{\bar{K}\pi} &=& a_2(m_D^2-m_\pi^2)f_Kf_0^{D\pi}(m^2_K),  \non \\
\a_{K\bar{\pi}} &=& a_1(m_D^2-m_\pi^2)f_Kf_0^{D\pi}(m^2_K),  \\
\b_{K\bar{\pi}} &=& a_2(m_D^2-m_\pi^2)f_Kf_0^{D\pi}(m^2_K), \non \\
\c_\kpi &=& \c_\bkpi. \non
\en
It is clear from Eq.(25) that $(\b)_{K\bar{\pi}}=(\b)_{\bar{K}\pi}$.
Therefore, the SU(3)-breaking effect in $R_2$ comes solely from the external
$W$-emission amplitude $\a$, which is $\a_{K\bar{\pi}}/\a_{\bar{K}\pi}
=1.487$, calculated from Eq.(25). The current model calculation is incapable of
 estimating the
$W$-exchange amplitude $\c$ since the relevant form factors have to be
evaluated at $q^2=m_D^2$. Using the model calculation of the amplitude $\a_
\bkpi$ given in Eq.(25) we can determine $\c$ from the measured
branching ratio of $D^0\to K^-\pi^+$ and $\tau(D^0)=4.20\times 10^{-13}s$ [6]
\be
\left({\c\over\a}\right)_{\bar{K}\pi}\cong -0.13\,.
\en
This shows that the $W$-exchange contribution is small but not negligible.
Putting everything together into Eq.(2), we find
\be
R_2=\,2.3\,.
\en
This is in agreement with the experimental result Eq.(6). (We would like to
comment that the
final value of $R_2$ is not very sensitive to the value of Eq.(26). For example
if we were to use  twice  of that value, we would obtain $R_2=2.6\,$.)

   To conclude, we have shown that the unexpected large
decay rates of the recently measured quark-mixing singly-suppressed decay
$D^+\to\pi^+\pi^0$ and the doubly-suppressed decay $D^0\to K^+\pi^-$ by CLEO,
hence the large ratios of $R_1$ and $R_2$,
can be accounted for in the large-$N_c$ factorization
approach. Accumulations   of small SU(3)-symmetry breaking effects in
the decay constants, form factors, and mass-difference ratio lead to
large values of $R_1$ and $R_2$, deviating from the SU(3)-symmetry values of
unity for both. An important ingredient  for our analysis to obtain the correct
values of $R_1$ and $R_2$  is the
relative magnitude of the form factors $f_+^{D\pi}(0)>f_+^{DK}(0)$, consistant
with the recent theoretical calculation [13] and a recent CLEO
measurement of $D\to\pi\ell\bar{\nu}$ [12].

\vskip 2.3 cm
\centerline{\bf Acknowledgments}
\vskip 1.0 cm
One of us (H.Y.C.) wishes to thank Professor C. N. Yang and the
Institute for Theoretical Physics at Stony Brook for their hospitality
during his stay there for sabbatical leave.  We would like to thank Professor
D. W. McKay for helpful comments.
This research was supported in part by the US Department of Energy and the
National Science Council of ROC under Contract No.  NSC83-0208-M001-014.

\vskip 4.0cm

\centerline{\bf References}
\medskip
\begin{enumerate}

\item L.L. Chau Wang and F.
Wilczek, \prl {\bf 43}, 816 (1979).

\item L.L. Chau, {\sl Phys. Rep.} {\bf 95}, 1 (1983).

\item L.L. Chau and H.Y. Cheng, \pr {\bf D36}, 137 (1987); \prl {\bf 56},
1655 (1986); {\bf A4}, 877 (1989); \pr {\bf D39}, 2788
(1989); \pl {\bf B222}, 285 (1989); \pr {\bf D42}, 1837 (1990).

\item CLEO Collaboration, M. Selen {\it et al.,} \prl {\bf 71}, 1973 (1993).

\item CLEO Collaboration, D. Cinabro {\it et al.,} \prl {\bf 72}, 406 (1994).

\item Particle Data Group, \pr {\bf D45}, S1 (1992).

\item Another long-standing problem of SU(3)-violating effects observed in
the
decays $D^0\to K^+ K^-$ and $D^0\to \pi^+\pi^-$ was discussed by B.F.L. Ward,
\prl {\bf 70}, 2533 (1993); A. Czarnecki, A.N. Kamal, and Q. Xu, \zp {\bf
C54}, 411 (1992); L.L. Chau and H.Y. Cheng, \pl {\bf B280}, 281 (1992).

\item H.Y. Cheng, {\sl Int. J. Mod. Phys.} {\bf A4}, 495
(1989).

\item The form factors $f_0$ and $f_1$ are related to the conventional ones
$f_+$ and $f_-$
through the relations $f_+(q^2)=f_1(q^2)$ and
\be
f_-^{D\pi}(q^2)=[(m^2_D-m^2_\pi)/ q^2][f_0^{D\pi}(q^2)-f_1^{D\pi}(q^2)]. \non
\en
Note that $f_0=f_1$ at $q^2=0$.

\item M. Witherell, talk presented at the XVI International Symposium
on Lepton-Photon Interactions, Ithaca, 10-15 August 1993.

\item Mark III Collaboration, J. Adler {\it et al.,} \prl {\bf 62}, 1821
(1989).

\item CELO Collaboration, M.S. Alam {\it et al.,} \prl {\bf 71}, 1311 (1993).

\item R. Casalbuoni, A. Deandrea, N. Di Bartolomeo, R. Gatto, F. Feruglio,
and G. Nardulli, \pl {\bf B299}, 139 (1993).

\item M. Wirbel, B. Stech, and M. Bauer, \zp {\bf C29}, 637 (1985).

\item L.L. Chau and H.Y. Cheng,  {\sl Mod. Phys. Lett.} {\bf A4}, 877 (1989).

\item D. Fakirov and B. Stech, {\sl Nucl. Phys.} {\bf B133}, 315 (1978);
B. Blok and M. Shifman, {\sl Sov. J. Nucl. Phys.} {\bf 45} (1987) p. 135,
301, 522.

\item A.J. Buras, J.-M. G\'erard, and R. R\"uckl, \np {\bf B268}, 16 (1986).
For early papers suggesting dropping the $1/N_c$ terms, see M. Fukugita,
T. Inami, N. Sakai, and S. Yazaki, \pl {\bf 72B}, 237 (1977); D. Tadic and
J. Trampetic, \pl {\bf 114B}, 179 (1982).

\item M.K. Gaillard and B.W. Lee, \prl {\bf 33}, 108 (1974); G. Altarelli
and L. Maiani, \pl {\bf 52B}, 351 (1974).

\item CLEO Collaboration, D.S. Akerib {\it et al.,} \prl {\bf 71}, 3070
(1993).

\end{enumerate}

\end{document}